# Hybrid-Rendering Techniques in GPU

Pedro Granja
pedro.granja@tecnico.ulisboa.pt

João Pereira
jap@inesc-id.pt

Instituto Superior Técnico, Lisboa, Portugal

July 2021

**Abstract**

Ray tracing has long been the holy grail of real time rendering. This technique, commonly used for photo realism, simulates the physical behavior of light, at the cost of being computationally heavy. With the introduction of Nvidia RTX graphic card family, which provides hardware support for ray tracing, this technique started to look like a reality for real time. However, the same problems that afflicted the usage of this technique remain, and even with specialized hardware it is still extremely expensive. To account for these drawbacks, researchers and developers pair this technique with rasterization and denoising. This results in a hybrid system that tries to join the best of both worlds, having both photo realistic quality and real time performance. In this work we intend on further exploring hybrid render systems, offering a **review of the state of the art with a special focus on real time ray tracing** and our own **hybrid implementation with photo realistic quality and real time performance** (>30 fps), implemented using the Vulkan API. In this project, we highlight the detailed analysis of the impacts of History Rectification (Variance Color Clamping) on the temporal filter component of the denoising system and how to overcome the introduced artifacts. Additionally, we also highlight the analysis of the introduction of a separable blur on the spatial filter and the introduction of Reinhard Tone Mapping prior to denoising, consequently improving this procedure.

## 1. Introduction

Real time ray tracing hardware disrupted the real time industry generating a lot of research on how to best use this technology. Initially, games adopted this technique scarcely, being used only for specific effects like ray traced shadows[1] or ray traced reflections[2]. With time, the adoption has been steadily increasing and RTX is slowly becoming its own rendering pipeline and used to compute part or full global illumination. This research culminated in games like Quake 2 with complete path traced global illumination.

With the advent of multiple implementations, a question remains... What is the best way to use this technique? And how much should we use it in our real time pipeline for quality while maintaining performance?

One popular solution, commonly used by the real time industry, is the combination of ray tracing and rasterization into a hybrid system. This technique tries to join the quality of ray tracing with the performance of rasterization. For this reason, in this work, we further explore this approach and offer our own implementation.

---
[1]Shadow of the Tomb Raider
[2]Battlefield V

## 2. Related Work

On most of the state of the art hybrid implementations the current major obstacle is denoising (Usually, denoising is applied over channels traced with only 1 sample per pixel (1spp) because of the cost associated with ray tracing). As such, a breakthrough was also required in this area and it came in the form of a conjunction of multiple techniques (temporal accumulation, spatial filtering, etc.) into a unified approach called Spatiotemporal Variance-Guided Filtering ([14]). This approach is capable of denoising channels with 1spp and when combined with ray tracing supported by hardware it is capable of presenting high quality, real time images. This technique is composed by 2 stages, a temporal filter with variance estimation and a spatial filter guided by variance.

The temporal filter uses backwards reprojection, allowing the current frame to be reprojected into the previous frame and as such, giving access to the previous temporal information in order to integrate the image results over time (essentially reducing the amount of noise present in the image). Furthermore, this technique also computes an estimation of variance, which provides a metric for the amount of noise present in each pixel.

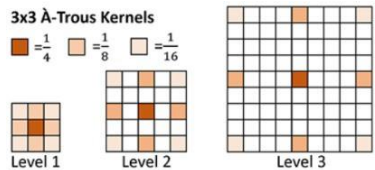

**Figure 1:** Representation of the differences in the kernel between 3 convolutions of the filter [17]

The spatial filter component of this method uses a technique called Edge-Avoiding À-Trous Wavelet Transform[5], which, similarly to a bilateral filter, samples the neighborhood information of the pixel while using edge preserving functions (estimate if the pixel being sampled is similar to the pixel being denoised. If the pixel pair is similar, than the information can be sampled). The major difference, comes in the form of the introduction of "holes" in the kernel of the blur over multiple iterations of the algorithm, reducing the computational cost (See Figure 1).

In Svgf, this technique is expanded in order to use the variance estimation calculated on the temporal filter. This expansion comes in the form of a new edge function which weights the luminance difference between pixels and the estimated variance for the pixel being denoised. Reducing the amount of information used when the variance is low (not a lot of noise present) and increasing when the variance is high (A lot of noise is present).

As a last step, temporal antialiasing is used, providing a final temporal pass for the algorithm and removing some additional noise.

This technique, however, still presents some artifacts, one of which due to temporal lag, noticeable on reflections with moving objects and detached shadows on fast moving geometry. This is caused, because, even though backwards reprojection accounts for geometry changes in the scene, it doesn't account for changes in the lighting effects (reflections, shadows, etc.) caused by this moving objects.

In order to solve this problem, the common solution is to extend Svgf temporal accumulation system with a technique that weights the differences in the colors of the pixels between the previous and the current frame. A possible solution for this problem is the usage of A-Svgf ([3]) which adds a temporal gradient calculated by reprojecting forward surface samples from the previous frame (Each pixel has a collection of attributes stored in the G-Buffer. It's this collection of attributes that the author denominates as surface samples). The temporal gradient is calculated by taking the difference between the shading on the previous frame and the current frame of the forward projected samples. This system has notably been used on Quake 2 RTX [15].

Another option is the usage of history rectification, a common technique used on temporal antialiasing ([12]). There are 2 approaches, color clamping and clipping. Color clamping and clipping start by calculating a color space bounding box based on a region surrounding the target pixel (ex: 3x3) on the current frame. This bounding box is than used to clip or clamp the previous frame pixel (See Figure 3), forcing the previous colors to fit the new ones. Color clipping and clamping are susceptible to outliers and as such were extended by a more robust variance version. Variance Color Clipping introduced by Salvi ([13]) estimates variance from the region surrounding the target pixel (ex: 3x3) and uses this information to create the color bounding box. This option has been used by systems like Pica Pica [4].

Another differentiation between Svgf and other systems being developed is the channels separation. While Svgf splits the illumination in a direct and indirect channel, and denoises each independently, the most common approach is to also separate the indirect specular channel and the indirect diffuse channel. This division allows different strategies to be employed for each effect.

Quake 2, even though it mostly uses A-Svgf, uses a simple Edge-Avoiding À-Trous Wavelet Transform guided by depth and normals, without variance computations for the indirect diffuse channel. This difference is justified by the sparsity of the noise produced in this channel, requiring more aggressive spatial techniques. In order to further blur over a larger surface, the authors use geometric normals instead of normal maps. Normal maps contain high frequency details that would reduce the amount of light contributions due to the edge avoidance component of the spatial filter. However, not using normal maps essentially means that the high frequency details of the illumination are lost. To combat this the authors, use spherical harmonics (particularly the 2 first spherical harmonics). Spherical harmonics allows the encoding of illumination combined with its directional information.

Another example of the separation of the indirect specular channel from the indirect diffuse channel is Battlfield V [11] and PICA PICA [4], where the indirect specular channel is ray traced in half resolution and the results are reconstructed in full resolution, assuming the neighboring pixels intersection results can be reused, resulting in a gain in performance. This is an approach commonly used on SSR (Stochastic Screen Space Reflections [16]). It is also common to separate the shadows from the direct illumination and denoise shadows inde-pendently. This is done, because, only the shadows computation rely on global scene information (Global objects positions need to be known because they can be occluding a light source) while direct illumination can easily be rasterized, even for area lights ([8], this has been expanded to join with stochastic shadows [9]). This also further reinforces the usefulness of an hybrid implementation, has the direct illumination can be rasterized [18].

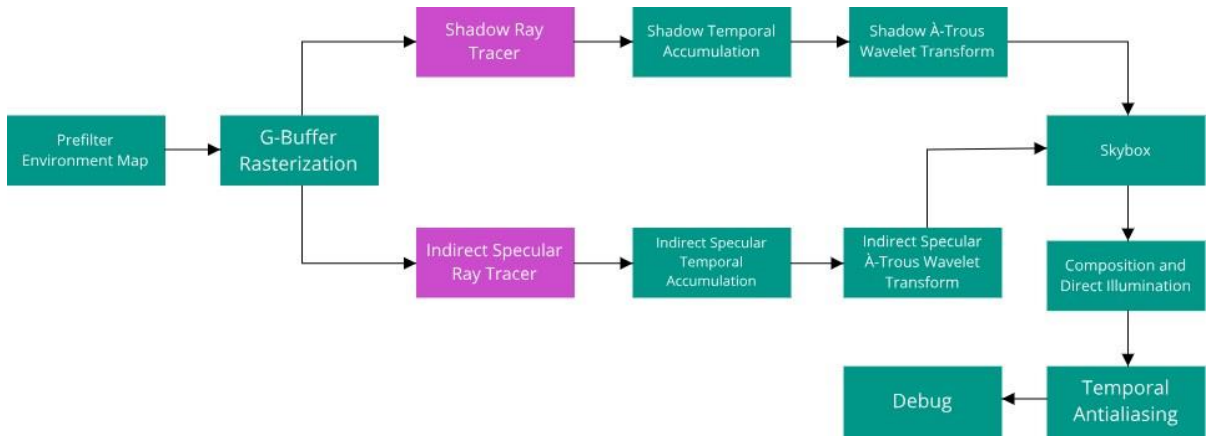

**Figure 2:** Application architecture.

**3. Implementation**

Our application structure and data flow is defined as follows:

1. Prefilter Environment Map - Skybox is prefiltered at the beginning of the application, as demanded by the image based lighting technique.

2. G-Buffer rasterization - Rasterize the necessary geometric information for the remaining passes.

3. Shadow Ray Tracer/Indirect Specular Ray tracer - Use the G-Buffer information to trace shadow rays/indirect specular rays.

4. Shadow/Indirect Specular Temporal Accumulation - Temporally accumulate the results from the ray tracers and calculate variance

5. Shadow/Indirect Specular À-Trous Wavelet Transform - Spatially denoise the temporally accumulated results using the variance has a guide.

6. Skybox - Render the background skybox

7. Composition and Direct Illumination - Blend the results from the denoisers and use a deferred rendering strategy in order to calculate direct illumination [7]

8. Temporal Antialiasing - Final temporal accumulation in order to remove some last remaining noise (We use the temporal antialiasing provided by Falcor [1]).

9. Debug - Final debug pass that allows the debugging of multiple buffers

The major key differences to Svgf are than as follows. Shadows are separated from direct illumination, guaranteeing that direct illumination is completely noise free. The indirect specular component is denoised separately from the indirect diffuse channel, allowing for custom denoising to be performed. However, due to time constraints, the indirect diffuse channel was not implemented. Both shadows and indirect specular channels go through a similar filter to Svgf with some extra techniques, these being: History rectification, Adapt Blur to Scene Features, Separable Blur and Reinhard Tone Mapping (Only applied to the indirect specular channel). We also implemented an image based lighting technique [6], which can be used for shading secondary ray bounces without introducing noise into the system (indirect specular channel).

**3.1. History Rectification**

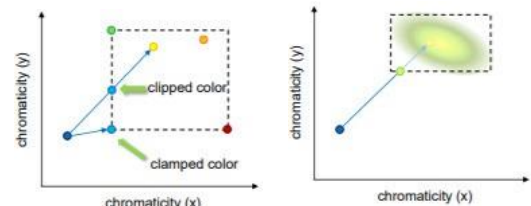

**Figure 3:** Example of color clipping and clamping on a 2D chromaticity space and variance color clipping. [12]

Two approaches where tested, variance color clipping and variance color clamping. The first step of both approaches is the calculation of the color variance and mean on the current frame pixel neighborhood (3x3). This values are than used to create a color "bounding box", which represents the values we expect the previous frame samples to conform too. Using these values, either clamping (variance color clamping) or clipping (variance color clipping) are used. By comparing both approaches, it became clear that there aren't a lot of differences between the two approaches on both shadows and reflections and as such we ultimately chose variance color clamping for our system. However, because this technique approximates the colors from the previous frame to the current frame it removes important ray tracing paths with low probability, causing temporal accumulation to become less efficient causing holes and jittering to become more noticeable and causing a divergence from the ground truth.

### 3.2. Adapt Blur to Scene Features

Due to history rectification reducing the efficiency of the temporal accumulation pass we opted by reinforcing the spatial filtering. Due to different surface roughnesses and shadow angles producing different degrees of noise (low surface roughnesses and shadow angles produce a reduced amount of noise and high surface roughnesses and high shadow angles produce a high amount of noise) we opted by starting the Edge-Avoiding À-Trous Wavelet Transform at a higher level of blur scale on certain thresholds (surface roughness over 0.2 and shadow angles over 6.0), having our first iteration be the equivalent to the usual second iteration. Because the number of iterations remain the same, this results in further pixels being visited and a more aggressive blur, reducing the holes and jittering. However, due to the increase in the blur levels, banding artifacts start becoming noticeable.

### 3.3. Separable Blur

Multiple blur iterations with a large number of neighbour visits are expensive. Has such, we separated the 5x5 blur kernels used in the usual Edge-Avoiding À-Trous Wavelet Transform into 2 separated passes, one vertical and one horizontal. This changes the number of neighborhood visits from 25(5x5) to 10 (5+5), representing considerable speed gains. In terms of quality, the results remain similar, although an increase in the amount of blur can be noticed, increasing some banding artifacts. This can also be a positive improvement in some regions with insufficient blur. The cause of the increase in blur is due to the variance only being updated in the second pass of the separable blur (vertical pass). Due to the positive improvement in regions with insufficient blurring we opted for continuing with this approach. This technique is mentioned on [2]

### 3.4. Reinhard Tone Mapping

Reinhard Tone Mapping is applied to the final results from the indirect specular ray tracer. This operation removes fireflies and reduces the divergence between multiple different pixels, creating a more "homogeneous" final ray tracing result. The inverse of this operation is than performed after denoising, however, it causes a divergence from the ground truth. A similar technique was used for Wolfenstein: Youngblood [10].

```
//Reinhard Tone Mapping before denoising
res /= (1.0f + luma(res) * lumaMultiplier);

//"Remove" reinhard tone mapping after
//denoising
specular = specular * (1.0f + luma(specular)) *
globalShaderInputs.reinhardToneMappingWeight;
```

### 3.5. Image Based Lighting

Low surface roughness should produce very dense noise and has such require a reduced amount of blurring. This is, however, not the case, has the subsequent ray bounces have no guarantees in regards to noise (The objects being reflected may be rough and has such produce sparse noise that requires a lot of spatial filtering in order to be denoised). Due to this property, we experimented with using an image based lighting technique in order to calculate the shading for the second ray intersections (Using this technique [6]), guaranteeing that no noise is present. This allows us to than change the number of blur iterations depending on the surface roughness, reducing over blurring on low surface roughnesses:

```
//Adapt the number of iterations
//to the amount of noise we expect
if(roughness <= 0.0) {
        maxBlurIteration = 0;
}
else if(roughness <= 0.05)
        maxBlurIteration = 1;
else
        maxBlurIteration = 4;
```

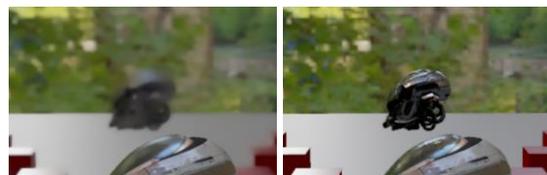

**Figure 4:** Comparison between not using Image Based Lighting for secondary ray intersections and using Image Based Lighting. Notice how in the second image the reflections aren't as blurred.

However, this technique is deactivated for the remaining of the document due to deviating the re-sults from the ground truth, nevertheless, it showcases how other rasterization techniques like lighting probes could be used in order to improve denoising, by guaranteeing a noise free second ray intersection. Another important point to mention is that this strategy also affects the denoising of shadows in reflections (high shadow angles visible through reflections produce a lot of noise) and has such would also need an alternative solution.

## 4. Results

We evaluate our application in both performance and quality (Using SSIM). We generalize the tests for both shadows and the indirect specular component, but this tests are performed separately in order to better see the impacts the techniques have in each. The parameters evaluated are thus the following:

- **Surface Roughness and shadow angle:** Shadow angle/Surface roughness of one or multiple materials.

- **Type of movement:** Our system is tested against multiple types of movement [Static (No movement), Moving camera, Moving Lights and Objects]. Due to presenting almost no differences to the static camera, the moving camera for shadows is not presented.

- **Denoising improvements:** Sequential improvements to our implementation of SVGF. The improvements showcased are the following [Our Svgf (baseline algorithm), Clipping, Blur adapted to surface roughness/shadow angle, Separable Blur, Reinhard Tone Mapping (In the indirect specular case)]. These techniques are built on top of one another and the results are progressively analysed.

For performance we follow a similar approach to the quality evaluation, comparing multiple surface roughnesses, shadow angles, types of movements and denoising improvements.

### 4.1. Quality Metrics
#### 4.1.1 Indirect Specular

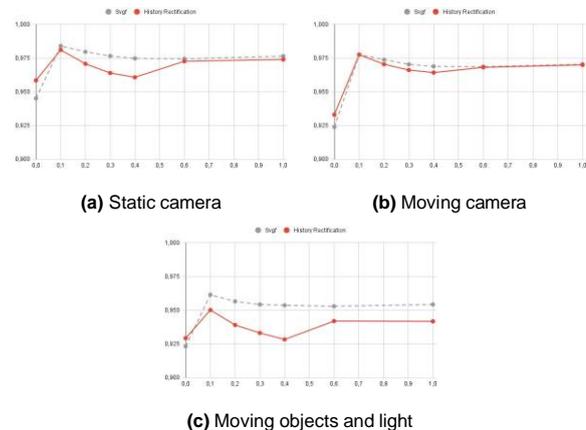

(a) Static camera  (b) Moving camera

(c) Moving objects and light

**Figure 5:** Impact of *History Rectification* in the Cubes Distance Scene (y axis represents the SSIM, x axis the surface roughness)

**History Rectification:**

- **Static:** The static scenes present an increase in quality at the 0.0 surface roughness values, a decrease in the mid tier surface roughness values ([0.2, 0.3, 0.4]) and 0.1 roughness values and, finally, a similar quality in the high tier surface roughnesses ([0.6, 1.0]). These results are to be expected given the removal of the low probability paths caused by history rectification. Temporally unstable noise becomes more apparent in the image (See Figure 18, cubes distance scene, static camera, close up green).

- **Moving Camera:** The results follow a similar pattern to the static tests, with an increase in quality in the low surface roughness values ([0.0, 0.1]), a decrease in the mid tier surface roughness values ([0.2, 0.3, 0.4]) and, finally, a similar result for the high tier surface roughnesses ([0.6, 1.0]). These similar (follow the same pattern) but better results can be explained by the improvements towards ghosting (See Figure 18, cubes distance scene, moving camera, close up green), which was not present in the static tests.

- **Moving Objects and Light:** A similar pattern to the previous graphs can be observed with but with a larger divergence from the ground truth at most surface roughenesses. Looking at both scenes, a lot of improvements can be seen towards ghosting artifacts (See Figure 18, cubes distance scene, moving objects and light, close up green).

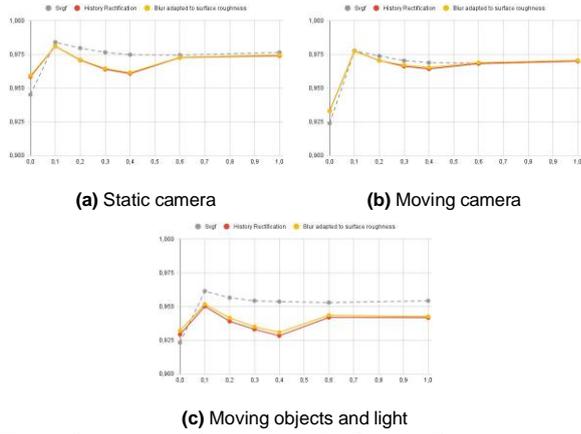

(a) Static camera  (b) Moving camera

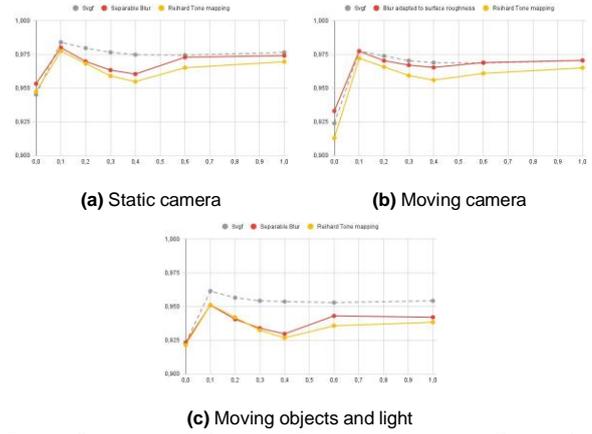

(a) Static camera  (b) Moving camera

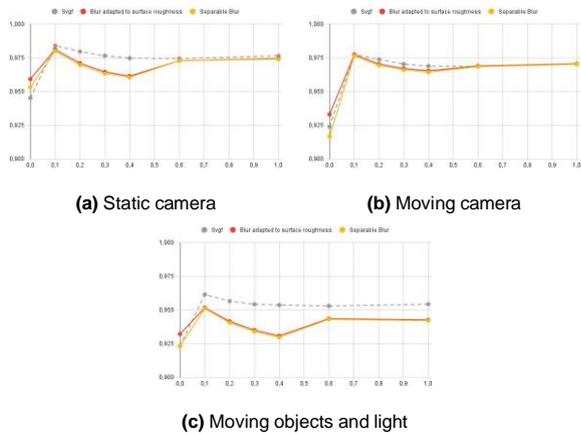

(c) Moving objects and light

**Figure 6:** Impact of *Blur Adapted To Surface Roughness* in the Cubes Distance Scene (y axis represents the SSIM, x axis the surface roughness)

(c) Moving objects and light

**Figure 8:** Impact of *Reinhard Tone Mapping* in the Cubes Distance Scene (y axis represents the SSIM, x axis the surface roughness)

**Blur Adapted To Surface Roughness:** As can be seen, the quality results from the introduction of this technique remain very similar to the previous entry, however, a reduction in the amount of temporally unstable noise can be noted (notice how in Figure 18, cubes distance scene, static camera, close up green, the results appear more homogenous after the introduction of this technique). Banding artifacts start to become visible (Figure 18, cubes distance scene, static camera, close up red).

**Reinhard Tone Mapping:** The introduction of this technique causes a slight divergence from the ground truth results, however, visually, the results stay largely the same.

### 4.1.2 Shadows

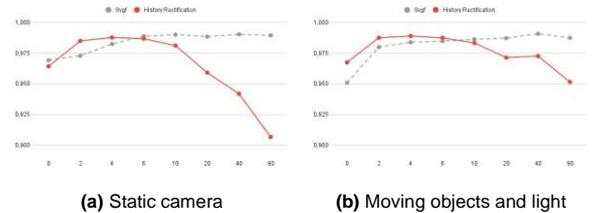

(a) Static camera  (b) Moving objects and light

**Figure 9:** Impact of *History Rectification* in the Breakfast Room Scene (y axis represents the SSIM, x axis the surface roughness)

**History Rectification:**

- **Static:** The quality metrics show improvements towards the low shadow angles, and a gradual worsening towards the high shadow angles. This worsening can be justified by the same reasoning as the reflections, the removal of low probability paths. Jagged lines and noise become more noticeable (See Figure 19).

- **Light and Object Movement:** The quality metrics show a similar trend to the results provided by the static camera, but more positive, as artifacts related to ghosting are removed (Figure 18, Breakfast Room scene, moving objects and light, close up red).

(a) Static camera  (b) Moving camera

(c) Moving objects and light

**Figure 7:** Impact of *Separable Blur* in the Cubes Distance Scene (y axis represents the SSIM, x axis the surface roughness)

**Separable Blur:** Similarly to the previous entry, the quality results stay largely unchanged with only the low surface roughnesses being affected due to over blurring (see the increase in blur in Figure 18 cubes distance scene, moving camera, close up green). Banding artifacts become more visible and can be easily noticeable (Figure 18, cubes distance scene, static camera, close up red).

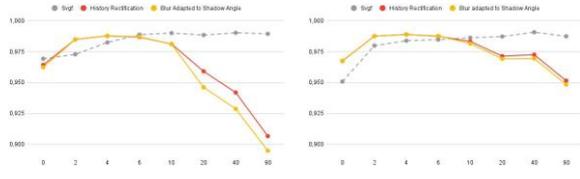

**(a)** Static camera  **(b)** Moving objects and light
**Figure 10:** Impact of *Blur Adapted to Shadow Angle* in the Breakfast Room Scene (y axis represents the SSIM, x axis the surface roughness)

**Blur Adapted To Shadow Angle:** As can be seen, the quality results from this technique remain very similar to the previous entry, however, a reduction in the amount of noise and jagged lines can be noted (See Figure 19).

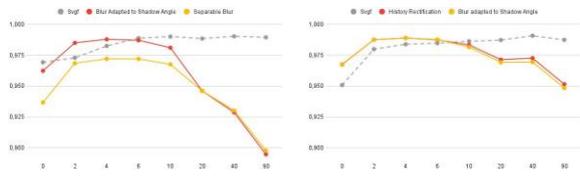

**(a)** Static camera  **(b)** Moving objects and light
**Figure 11:** Impact of *Separable Blur* in the Breakfast Room Scene (y axis represents the SSIM, x axis the surface roughness)

**Separable Blur:** Similar to the previous entry, the quality results stay similar with only the low shadow angles being affected due to over blurring.

**4.2. Performance Metrics**

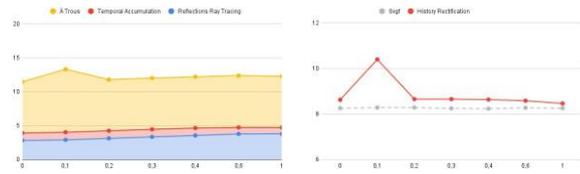

**(a)** Main passes time comparison  **(b)** Techniques comparison
**Figure 12:** Indirect Specular - *History rectification* impact on performance in the Cubes Distance Scene

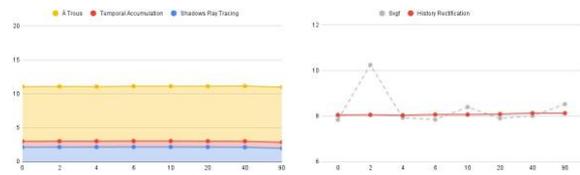

**(a)** Main passes time comparison  **(b)** Techniques comparison
**Figure 13:** Shadows - *History rectification* impact on performance in the Breakfast Room Scene

**History Rectification:** Has can be expected, this technique slightly increases the timing in both shadows and the indirect specular channel. A few interesting observations can also be made. Firstly, all the blur iterations of the À Trous filter take a similar time (For Svgf on the Cubes Distance scene with 0.3 surface roughness, [1.93ms, 1.87ms, 1.86ms, 1.86ms]). This is surprising due to À Trous visiting progressively further away pixels, and as such, due to data locality, it was expected to increase the timings. Another important observation, is the impact of the surface roughness on the time it takes the ray tracing of the indirect specular channel. Has can be observed, there is an increase in time proportional to the surface roughness. This is, most likely, a consequence of data locality. Has the surface roughness increases, the rays have wider differences in directions and hit points, being more divergent between them. A way to reduce the impact of this is to use a ray binning strategy.

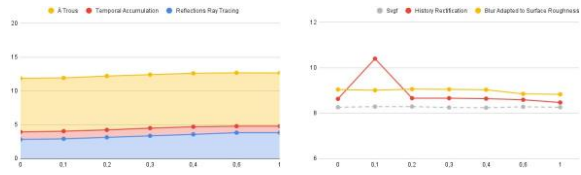

**(a)** Main passes time comparison  **(b)** Techniques comparison
**Figure 14:** Indirect Specular - *Blur adapted to surface roughness* impact on performance in the Cubes Distance Scene

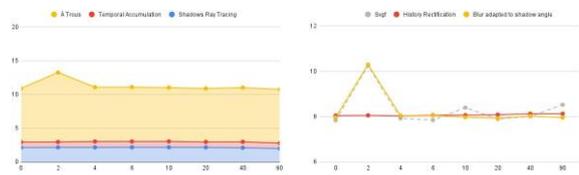

**(a)** Main passes time comparison  **(b)** Techniques comparison
**Figure 15:** Shadows - *Blur adapted to shadow angle* impact on performance in the Breakfast Room Scene

**Blur Adapted to Shadow Angle/Surface Roughness:** The shadow results stay similar to the previous iteration of the algorithm. A slight decrease in performance can be noticed on the indirect specular results. The exact reasoning for this increase in performance could not be found. One might expect data locality to be the cause, but the low surface roughnesses seem to also be affected by this delay (low surface roughnesses have a reduced blur size).

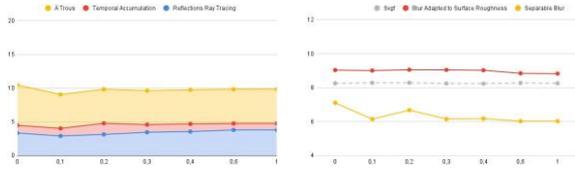

**(a)** Main passes time comparison    **(b)** Techniques comparison
**Figure 16:** Indirect Specular - *Separable Blur* impact on performance in the Cubes Distance Scene

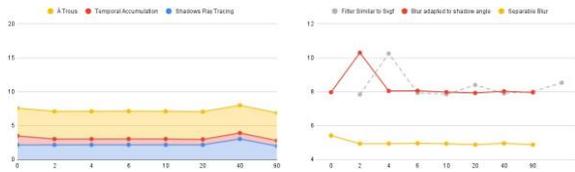

**(a)** Main passes time comparison    **(b)** Techniques comparison
**Figure 17:** Shadows - *Separable Blur* impact on performance in the Breakfast Room Scene

**Separable Blur:** Has can be observed, this technique greatly improves the performance of the spatial procedure.

### 4.2.1 Animation performance:

| Timings | Cubes Distance | Icosphere | Glossy Sponza 1 | Glossy Sponza 2 |
|---|---|---|---|---|
| max | 80,1 | 79,8 | 51,9 | 45,6 |
| avg | 79,2 | 77,2 | 51,5 | 45,3 |
| min | 76,5 | 76,1 | 51,2 | 45,2 |

**Table 1:** Performance ranges for animated sequences on multiple different scenes

| Timings | Glossy Sponza 3 | Shadow Objects | Pillars | Breakfast Room |
|---|---|---|---|---|
| max | 44,5 | 86,8 | 83,7 | 45,5 |
| avg | 43,4 | 84,4 | 82,6 | 45,1 |
| min | 43,1 | 83,8 | 82,1 | 44,9 |

**Table 2:** Performance ranges for animated sequences on multiple different scenes (Continuation)

Looking at table 1 and 2 it is possible to see that our final results stay well over the established goal of > 30fps, even on more complex scenes containing a lot of geometry and materials. However, the introduction of the indirect diffuse component would probably be too costly for achieving our goals.

### 5. Conclusions

This project ended up being a very challenging yet rewarding experience, requiring a lot of different areas and involving a multitude of disciplines (rendering techniques, shading and image processing).

Given our initial proposal of "creating a hybrid implementation with photo realistic quality and real time performance (> 30 fps)" we have achieved most of our goals with exception to the indirect diffuse component of lighting due to time constraints.

We also introduce a set of techniques over the original Svgf, allowing the algorithm to perform better in motion and have an increase in quality (SSIM) in some cases (low surface roughness and low shadow angle). The improvements in motion are due to History Rectification (Variance Color Clamping), resulting in reduced ghosting artifacts and a better response to movement and changes in the scene. This technique, however, reduces the efficiency of temporal accumulation, and as such introduces more temporally unstable noise into the scene.

To counteract this problem, the initial blur size is adapted to the features of each lighting component (surface roughness and shadow angle), increasing when a large amount of noise is expected (high surface roughness and high shadow angle). This also further shows the importance of separating the denoising procedure into multiple channels per lighting component.

Reinhard Tone Mapping is also shown to be a valuable way of improving denoising on the indirect specular component of lighting, trading ground truth correctness for less perceived flickering. Blurring operations over multiple iterations are a very expensive operation and as such we introduced a separable blur, where the normal 5x5 blur kernel is separated into a vertical and horizontal pass, greatly increasing the performance of the algorithm. However, artifacts like banding become more prevalent and the results become sightlier blurred (reducing some artifacts related to insufficient blurring but also over blurring some regions).

It is also shown, that the introduction of rasterization techniques (Image Based Lighting) can be introduced on second light bounces, allowing for a reduction on aggressiveness in denoising on objects with low surface roughness (Due to the guarantee that tertiary rays are noise free), resulting in a cleaner and less overblurred result. Although our results use an inaccurate version of IBL, this method could easily be replaced by precomputed lighting probes.

## 6. Image Results

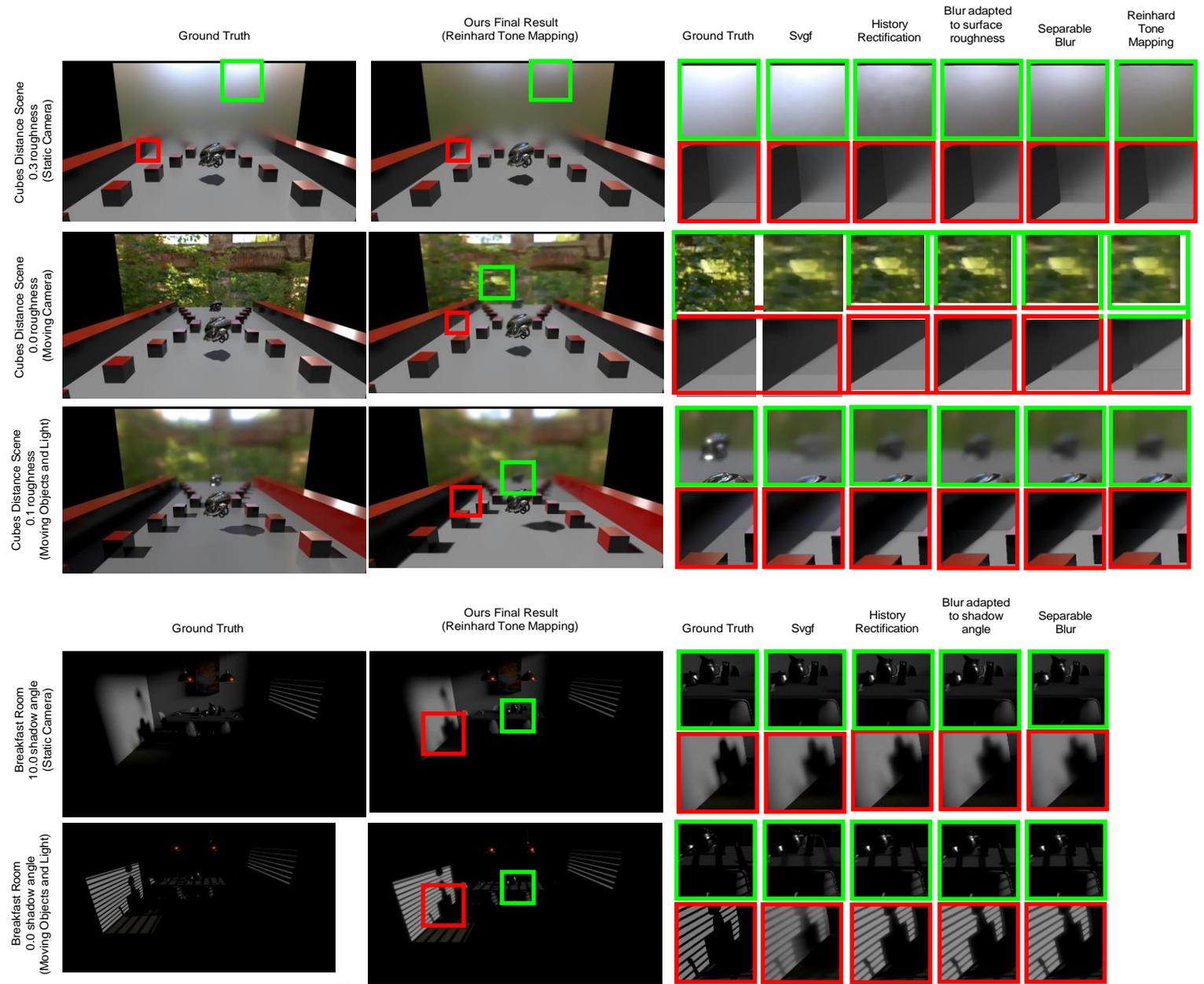

**Figure 18:** Comparison between multiple techniques and movement states

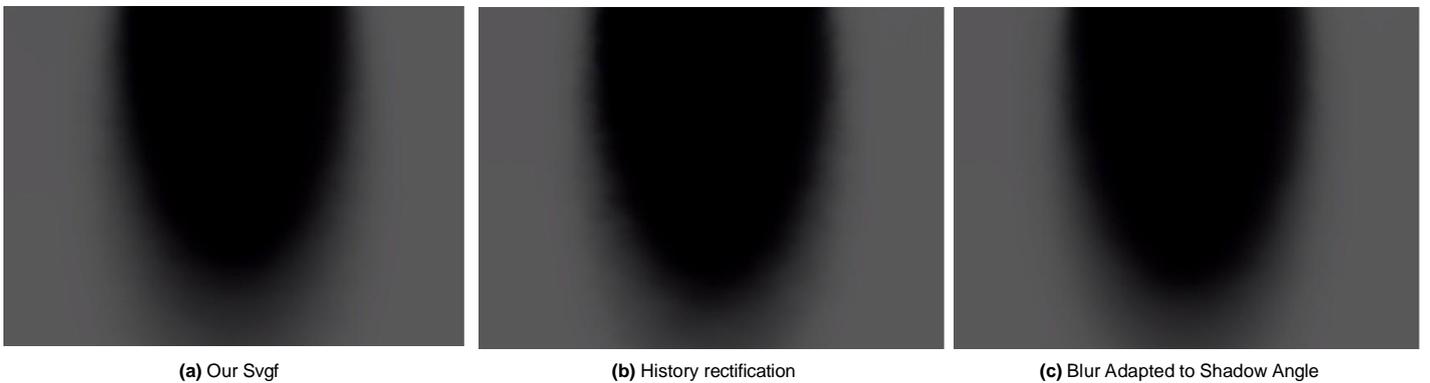

**Figure 19:** Comparison between Our Svgf, History rectification and Blur adapted to Shadow Angle on Shadow Objects scene with shadow angle with value 10.0. Images have increased contrast in order to better notice the jagged lines